\documentclass[12pt]{article}
\usepackage{graphicx}
\usepackage{epsfig}
\def\bs{\vspace{5pt}}

\newcommand{\Z}{{\widehat Z}}
\newcommand{\hH}{{\widehat H}}
\newcommand{\re}{{\hat r}_{\rm e}}

\newcommand{\tZ}{{H}}
\newcommand{\beq}{\begin{equation}}
\newcommand{\eeq}{\end{equation}}
\newcommand{\beqa}{\begin{eqnarray}}
\newcommand{\eeqa}{\end{eqnarray}}
\newcommand{\beqar}{\begin{eqnarray*}}
\newcommand{\eeqar}{\end{eqnarray*}}

\newcommand{\labell}[1]{\label{#1}} 
\newcommand{\reef}[1]{(\ref{#1})}

\newcommand{\eg}{{\it e.g.,}\ }
\newcommand{\ie}{{\it i.e.,}\ }

\newcommand{\norm}[1]{\raise.3ex\hbox{:}#1\raise.3ex\hbox{:}}

\newcommand\prt{\partial}

\def\simge{
    \mathrel{\rlap{\raise 0.511ex 
        \hbox{$>$}}{\lower 0.511ex \hbox{$\sim$}}}}

\def\simle{
    \mathrel{\rlap{\raise 0.511ex 
        \hbox{$<$}}{\lower 0.511ex \hbox{$\sim$}}}}

\begin{document}

\setlength{\unitlength}{1mm}

\thispagestyle{empty}
\rightline{\small hep-th/0105077  \hfill McGill/01-03}
\rightline{\small \hfill UTPT-01-07}
\rightline{\small\hfill  NSF-ITP-01-21}
\rightline{\small \hfill DCTP--01/35}
\vspace*{1.0cm}

\begin{center}
{\bf \Large The Enhan\c con and the Consistency of Excision}\\
\vspace*{1cm}

{\bf Clifford V. Johnson$^{a,}$\footnote{e--mail: {\tt
      c.v.johnson@durham.ac.uk}}, Robert C.
    Myers$^{b,}$\footnote{e--mail:  {\tt rcm@hep.physics.mcgill.ca}}, Amanda
    W. Peet$^{c,}$\footnote{e--mail:  {\tt peet@physics.utoronto.ca}}, Simon
    F. Ross$^{a,}$\footnote{e--mail:   {\tt s.f.ross@durham.ac.uk}} }

\vspace*{0.5cm}

${}^{a}${\it Centre for Particle Theory}\\
{\it Department of Mathematical Sciences, University of Durham}\\
{\it Durham DH1 3LE England, U.K.}\\[.5em]

${}^b${\it Department of Physics, McGill University}\\
{\it Montr\'eal, Qu\'ebec, H3A 2T8, Canada}\\[.5em]

${}^c${\it Department of Physics, University of Toronto}\\
{\it 60 St. George St., Toronto Ontario, M5S 1A7, Canada}\\[.5em]

\vspace{1.5cm} {\bf Abstract}
\end{center}
The enhan\c con mechanism removes a family of time--like singularities
from certain supergravity spacetimes by forming a shell of branes on
which the exterior geometry terminates. The problematic interior
geometry is replaced by a new spacetime, which in the prototype
extremal case is simply flat. We show that this excision process, made
inevitable by stringy phenomena such as enhanced gauge symmetry and
the vanishing of certain D--branes' tension at the shell, is also
consistent at the purely gravitational level. The source introduced
at the excision surface between the interior and exterior geometries 
behaves exactly  as a shell of wrapped D6--branes, and in particular,
the tension vanishes at precisely the enhan\c con radius.
These observations can be generalised, and we present the
case for non--extremal generalisations of the geometry, showing
that the procedure allows for the possibility that the interior
geometry contains an horizon. Further knowledge of the dynamics of the
enhan\c con shell itself is needed to determine the precise position
of the horizon, and to uncover a complete physical interpretation of the
solutions.

\vfill \setcounter{page}{0} \setcounter{footnote}{0}
\newpage

\section{Introduction}

In ref.~\cite{jpp}, the study of the supergravity fields produced by a
family of brane configurations revealed a new mechanism by which
string theory removes a class of time--like naked singularities. The
singularities, of ``repulson'' type, arise at the locus of points
where parts of the ten dimensional geometry shrink to zero size.  The
supergravity geometries preserve eight supercharges. In the prototype
example of ref.~\cite{jpp} it is a K3 manifold (on which the branes are
wrapped) which shrinks to zero size, but the presence of a K3 is not
essential for the phenomenon.

In short, the na\"\i ve supergravity solution is modified by the fact that
the constituent branes which source the fields smear out from being
point--like (in their transverse space) to being a sphere.  This sphere
is called the ``enhan\c con''. Pure supergravity is unable to model
this phenomenon, because it is controlled by physics which arises
before the shrinking parts of the geometry get to zero size. Instead,
when they get to volumes set by a characteristic length
$2\pi\sqrt{\alpha^\prime}$, new massless modes appear in the string
theory. In ref.~\cite{jpp}, the shrinking  volume of K3 gets to
$V_*=(2\pi\sqrt{\alpha^\prime})^4$, and there is an enhanced gauge
symmetry $U(1)\to SU(2)$. The unwrapped parts of the D--branes are
monopoles of the $U(1)$ and correspondingly become massless and expand
at this place, forming the enhan\c con.

The supergravity geometry interior to the enhan\c con, containing the
repulson singularity, is obviously incorrect, as it exhibits a number
of unphysical properties uncovered in, for example,
refs.~\cite{jpp,rep1,rep2,rep3}.  In fact, since it has a naked
singularity, the repulson is not only unphysical, it is part of a
family of incorrect geometries with the correct asymptotic charges,
since no--hair theorems (which are usually relied upon, at least
implicitly, to interpret supergravity physics) apply only if a
singularity in the proposed geometry is hidden behind an horizon.  The
proposal~\cite{jpp} was therefore that the repulson be excised and
replaced with a more appropriate geometry. In the cases studied in
ref.~\cite{jpp}, the geometry in the interior is simply flat space,
since there are no brane sources in the interior, as they have all
expanded out to form the enhan\c con.

In section 2 of this paper, we will show that this excision procedure
is consistent ---and in fact is an extremely natural process--- in
supergravity. By analysing the standard junction conditions, we show
in the next section that the enhan\c con radius is a special place
even from the simple point of view of the stress--energy of the shell,
and that this stress--energy corresponds precisely to that of a shell
of wrapped D6--branes.  We also show that the shell provides sources
for the dilaton and R--R fields which again match precisely to those
of wrapped D6--branes.

In section \ref{d2} we generalise this situation slightly by adding
D2--branes, to set the stage for section \ref{gooddef}, where we study
two families of non--extremal generalisations of the enhan\c con
geometry.  The first family corresponds to a system combining wrapped
D6--branes and a larger number of additional D2--branes.  These
non--extremal solutions all contain an event horizon, which may or may
not appear outside of the enhan\c con radius.  In the case where an
event horizon appears below the enhan\c con radius, the excision
procedure gives us a range of choices for the interior solution.  It
does not appear that this ambiguity can be resolved within the
supergravity framework alone.

The second family, which is characterised by a repulson--like
singularity (before any excision) is an extension of those derived in
ref.~\cite{jpp}\footnote{We correct a small but crucial typographical
  error in the non--extremal solution presented in ref.~\cite{jpp}.}.
This solution seems to describe a non--extremal configuration with
arbitrary numbers of D2-- and D6--branes. However, this solution does
not reduce to the previous one when the number of D2--branes exceeds
the number of D6--branes, nor does it reduce to a standard D6--brane
solution when the volume of the $K3$ is large. Furthermore, this
solution also has the peculiar feature that it never has an event
horizon outside of the enhan\c con radius. Hence its physical
interpretation remains unclear. Section \ref{concl} presents
conclusions and discussion of future directions.

Although perhaps in retrospect it could not have been much different,
we do find it remarkable that while supergravity cannot produce the
stringy phenomena which make the enhan\c con mechanism necessary, it
does display some awareness of the behaviour of the branes that source
this geometry; the source terms on the shell correspond to precisely
those in the worldvolume action.

\section{The Extremal Enhan\c con}
\label{extremal}

The enhan\c con story is best told by considering the case of wrapping
$N$ D$(p+4)$--branes on a K3 manifold of volume $V$. This leaves an
unwrapped $(p+1)$--dimensional worldvolume in the non--compact six
dimensions. There are $5-p$ non--compact spatial dimensions transverse
to the brane. We shall often use polar coordinates in these
directions, since everything we do here will retain rotational
symmetry. These coordinates are $r,\{\Omega_{4-p}\}$, where the set
$\{\Omega_{4-p}\}$ denotes one's favourite choice of angular 
coordinates on a unit round $(4-p)$--sphere, $S^{4-p}$.

The supergravity solution necessarily arranges that the volume of K3
decreases from the value $V$ at $r=\infty$ to smaller values as $r$
decreases. We shall denote this running volume as~$V(r)$.  Type II
string theory compactified on K3 has an enhanced gauge symmetry when
the K3 volume reaches $V_*=(2\pi)^4(\alpha^\prime)^2$, in our units.
The special radius at which this happens is called the enhan\c con
radius and denoted $r_{\rm e}$. We give its value below.

\subsection{The Geometry}

To avoid unnecessary notational clutter we focus on the case
$p=2$. For the issues that we consider, the extension to other $p$ is
trivial, and so we suppress those cases in the discussion, but the
reader may wish to keep them in mind for their own purposes. The
repulson solution is then
\begin{eqnarray}
ds^2 &=& Z_2^{-1/2} Z_6^{-1/2} \eta_{\mu\nu} dx^\mu dx^\nu +
Z_2^{1/2} Z_6^{1/2} dx^i dx^i + V^{1/2} Z_2^{1/2} Z_6^{-1/2} ds^2_{\rm K3} \
,\nonumber
\\
e^{2\Phi } &=& g_s^2 {Z_2}^{1/2}{ Z_6}^{-3/2}\ , \nonumber\\
C_{(3)} &=& ({Z_2} g_s)^{-1} dx^0 \wedge dx^1 \wedge dx^2\ , \nonumber\\
C_{(7)} &=& ({Z_6} g_s)^{-1} dx^0 \wedge dx^1 \wedge dx^2 \wedge
V\,\varepsilon_{\rm K3}
\ , \label{outside}
\end{eqnarray}
where the above line element corresponds to the string frame metric.
Here indices $(\mu,\nu)$ run over the 012--directions along the
unwrapped worldvolume, while indices $(i,j)$ run over the
345--directions transverse to the brane. Also,  $ds^2_{\rm K3}$ is the
metric of a K3 surface of unit volume and $\varepsilon_{\rm K3}$ is the
corresponding volume form. The harmonic functions are 
\begin{eqnarray}
Z_6 &=& 1+{r_6\over r}\ ,\quad Z_2 =1 -{r_2\over r}\ ,
\label{harms}
\end{eqnarray}
where
\begin{eqnarray}
\quad r_6 = {g_sN\alpha'^{1/2}\over 2} \ ,\quad r_2={V_*\over V} {r_6}\ ,
\label{arrs}
\end{eqnarray}
with $N$ being the number of D6--branes.
The running K3 volume can be simply read off as
\begin{equation}
V(r)=V{Z_2(r)\over Z_6(r)}\ .
\end{equation}
Using this, a  quick computation shows that 
\begin{equation}
 r_{\rm e} = {2V_*\over {V} - V_* } r_6\ .
\label{theradius}
\end{equation}
It is interesting to note that the following is true:
\begin{equation}
 {\partial\over\partial
r}\left(Z_2(r)Z_6(r)\right)\biggr|_{r=r_{\rm e}}=0\ .
\label{crucial}
\end{equation}
This simple result is in fact at the heart of the consistency of the
full junction computation we present in the next section, as we shall
see.

\subsection{Brane Probes}

The wrapped D6--branes will expand into a shell of zero tension at $r=r_{\rm
e}$. Just to remind the reader, and to set up the notation for the
following sections, we reproduce below the probe
computation of ref.~\cite{jpp} which supports this conclusion.

The effective worldvolume action of a single wrapped D6--brane, with
which we can probe the geometry, is:
\begin{equation}
S = - \int_{{\cal M}_2}d^3\xi\, e^{-\Phi(r)} (\mu_6 V(r) - \mu_2)
(-\det{G_{\mu\nu}})^{1/2} + \mu_6 \int_{{\cal M}_2 \times\rm K3}\! C_{(7)}
- \mu_2
\int_{{\cal M}_2} C_{(3)}\ , 
\label{probeaction}
\end{equation}
where ${\cal M}_2$ is the unwrapped part of the worldvolume, which lies
in six non--compact dimensions, and $G_{\mu\nu}$ is the induced (string frame)
metric. Of course, this result includes the
subtraction of one fundamental unit of D2--brane
tension\cite{bbg} and R--R charge\cite{bsv,anom,sunil}, which results
from wrapping the D6--brane on K3.
Note the R--R charges $\mu_p$ appear in the Dirac--Born--Infeld part of the
action since we have included the string coupling $g_s$ in the
solution for the dilaton. Recall that the basic D--brane tension 
is given by $\tau_p=\mu_p/g_s$.  The fundamental D6- and D2--brane charges
are $\mu_6 = (2\pi)^{-6}
\alpha'^{-7/2}$ and $\mu_2 = (2\pi)^{-2} \alpha'^{-3/2}$. Note that
\begin{equation}
{\mu_2\over\mu_6}={\mu_{p}\over\mu_{p+4}}=(2\pi)^4(\alpha^\prime)^2=V_*\ .
\end{equation}
It is the fact that this ratio yields $V_*$ for all $p$, 
following from T--duality, which
is at the heart of the consistency of the whole mechanism. If it
were not true, wrapped D4--branes would become massless at a different
value of $r$ from where wrapped D6--branes go massless, and then there
would be no W--bosons to carry the enhanced gauge symmetry. This
universality also underlies why we can focus on the case $p=2$ without
loss of generality.

Choose a static gauge where we align the worldvolume coordinates
$\xi^\mu$ with the first three spacetime coordinates $(x^0, x^1,
x^2)$, and then allow the transverse location of the brane to depend
only on time $t\equiv x^0$. Next, substitute the exterior solution
(\ref{outside}) into the action (\ref{probeaction}), and expand it,
keeping only quadratic order in velocity in $x^i$. In this way, one
can write the effective Lagrangian density for the problem of moving
the probe brane slowly in the background produced by all of the other
branes, ${\cal L}=T(r,\theta,\phi)-U(r)$. As this is a supersymmetric
problem, $U(r)$ is constant (it is zero in our conventions), while for
the kinetic energy we have:
\begin{eqnarray}
{T}&\equiv&{1\over 2}\tau(r) v^2={1\over 2g_s}
(\mu_6 V Z_2 -
\mu_2 Z_6) v^2  \nonumber\\
&=&{\mu_2 Z_6(r)\over 2g_s}\left[ {V(r)\over V_*}-1\right]
[{\dot r}^2+r^2({\dot \theta}^2+\sin^2\theta{\dot\phi}^2)]\ .
\label{kinetic}
\end{eqnarray}
Note that the second line shows
that the kinetic energy (and hence the effective tension $\tau(r)$ of
the probe) vanishes precisely at $r=r_{\rm e}$.

If we were to continue the result into the repulson region $r<r_{\rm
  e}$, we would find that the tension of the probe becomes negative.
Fixing that problem by taking the absolute value of the tension then
produces another problem: the cancellation of the potential $U(r)$
will fail, which is inconsistent with the supersymmetry of the
situation. Therefore the probe cannot proceed any further inside the
geometry than the enhan\c con radius.  We should also point out here
that more can be deduced~\cite{jpp,fuzz2,fuzzy} about the geometry by
exploiting the fact that the brane is in fact a BPS monopole of the
six dimensional Higgsed $SU(2)$. So not only does its mass go to zero
at the enhanced gauge symmetry point, but its size diverges, and the
probe spreads all over the enhan\c con locus.  The enhan\c con
therefore is a shell of smeared branes of zero tension.

 Since there are no point--like sources inside, it is not unreasonable
to suggest that the interior of the geometry is in fact flat space (in
this large $N$ approximation; there will be subleading corrections),
and we shall next examine the consistency of this proposal from the
point of view of supergravity.

\subsection{The Junction Conditions}

The brane probe calculation suggests that the repulson geometry is
replaced by flat space inside a shell, which is produced by
delocalised branes. We can use the classic gravitational
techniques~\cite{junction,misner} to describe this geometry more
explicitly, and calculate the stress--energy and charges of the shell
matching the exterior repulson geometry to flat space.

If we join two solutions across some surface, there will be a
discontinuity in the extrinsic curvature at the surface. This can be
interpreted as a $\delta$--function source of stress--energy located at
the surface. In the following section, we will show that the value of
this source precisely agrees with the source inferred
from the D--brane worldvolume action, confirming the consistency of this
description. 

Let us compute the relevant quantities, working at arbitrary incision
radius $r=r_{\rm i}$. The computation should be performed in Einstein
frame, to allow us to interpret the discontinuity in the extrinsic
curvature as a stress--energy~\cite{junction,misner}. One gets the ten
dimensional Einstein metric from the string one presented in the
previous section by the conformal rescaling
\begin{equation}
ds^2_{\rm E}= e^{-\Phi/2} ds^2_{\rm S}\ ,
\end{equation}
and we shall denote the metric components simply as $G_{AB}$, with no
further adornment. When we need to refer to  string frame metric
components, we shall be very explicit.

We are slicing in a direction perpendicular to the coordinate $r$, and
so we can define unit normal vectors:
\begin{equation}
n_\pm=\mp{1\over\sqrt{G_{rr}}}\,{\partial\over\partial r}\ ,
\end{equation}
where $n_+$ ($n_-$) is the outward pointing normal for the 
spacetime region $r>r_{\rm i}$ ($r<r_{\rm i}$).
In terms of these, the extrinsic curvature of the junction surface
for each region is
\begin{equation}
 K^\pm_{AB} = {1\over2}\,n_\pm^C\prt_C G_{AB}
=\mp{1\over 2\sqrt{G_{rr}}}\,{\partial G_{AB}\over\partial r} \ .
\label{extrinsic} 
\end{equation}
The discontinuity in the extrinsic curvature across the junction is
defined as $\gamma_{AB} =K^+_{AB}+K^-_{AB}$, and with these
definitions, the stress--energy tensor supported at the junction is
simply
\begin{equation} S_{AB}={1\over \kappa^2}
\left(\gamma_{AB}-G_{AB}\,\gamma^C{}_C\right)\ ,
\label{stresstensor} 
\end{equation}
where $\kappa$ is the gravitational coupling, \ie
$2\kappa^2=16\pi G_N=(2\pi)^7(\alpha^\prime)^4g_s^2$ is the
ten dimensional Newton's constant in our present conventions.

In our case, we want to match the Einstein metric of the repulson, 
\begin{eqnarray}
g_s^{1/2}\,ds^2 &=& Z_2^{-5/8}
Z_6^{-1/8} \eta_{\mu\nu} dx^\mu dx^\nu + Z_2^{3/8} Z_6^{7/8} dx^i dx^i
+ V^{1/2} Z_2^{3/8} Z_6^{-1/8}ds^2_{K3}\nonumber\\
&=& G_{\mu\nu}dx^\mu dx^\nu+G_{ij}dx^idx^j+ G_{ab}dx^adx^b \ ,
\label{outtside}
\end{eqnarray}
where $Z_2$ and $Z_6$ are given by (\ref{harms}), to a flat metric. We
will start with a matching at some arbitrary radius $r=r_{\rm i}$, and
will see that $r=r_{\rm e}$ is a special choice. We explicitly ensure
that all fields are continuous through the incision by writing the
interior solution in appropriate coordinates and gauge,
\begin{eqnarray}
g_s^{1/2}ds^2 &=& Z_2(r_{\rm i})^{-5/8} Z_6(r_{\rm i})^{-1/8}
\eta_{\mu\nu} dx^\mu 
dx^\nu + Z_2(r_{\rm i})^{3/8} Z_6(r_{\rm i})^{7/8} dx^i dx^i
\nonumber\\
&&\qquad\qquad\qquad\qquad\qquad\qquad
+ V^{1/2} Z_2(r_{\rm i})^{3/8} Z_6(r_{\rm i})^{-1/8} ds^2_{\rm K3} \
,\nonumber 
\\
e^{2\Phi} &=& g_s^2 Z_2^{1/2}(r_{\rm i}) Z_6^{-3/2}(r_{\rm i})\ , \nonumber\\
C_{(3)} &=& ({Z_2(r_{\rm i})} g_s)^{-1} dx^0 \wedge dx^1 \wedge dx^2\ ,
\nonumber\\
C_{(7)} &=& ({Z_6}(r_{\rm i}) g_s)^{-1} dx^0 \wedge dx^1 \wedge dx^2 \wedge
V\,\varepsilon_{\rm K3}
\ . \label{inside}
\end{eqnarray}
Some computation gives the following  results for the discontinuity tensor:
\begin{eqnarray}
\gamma_{\mu\nu}&=& \phantom{-}
{1\over 16}{1\over\sqrt{G_{rr}}}
\left(5{Z_2^\prime\over Z_2}+{Z_6^\prime\over Z_6} \right) 
G_{\mu\nu}\ , \nonumber \\
\gamma_{ij}&=& 
{-}{1\over 16}{1\over\sqrt{G_{rr}}}
\left(3{Z_2^\prime\over Z_2}+7{Z_6^\prime\over Z_6} \right) 
  G_{ij}\ , \nonumber \\
\gamma_{ab}&=& 
{-}{1\over 16}{1\over\sqrt{G_{rr}}}
\left(3{Z_2^\prime\over Z_2}-{Z_6^\prime\over Z_6} \right) 
  G_{ab}\ ,
\end{eqnarray}
where the metric components are as defined in eqn.~(\ref{outtside}),
a prime denotes $\partial_r$ and all quantities are evaluated at the incision
surface $r=r_{\rm i}$. Note, however, that $\gamma_{AB}$ is a tensor on
the nine dimensional junction surface and so $G_{ij}$ denotes the metric
on the angular directions of the transverse space, \ie there is no $\gamma_{rr}$
component. From the above we can compute the trace
\begin{equation}
\gamma^C{}_C={-}{1\over 16}{1\over\sqrt{G_{rr}}}
\left(3{Z_2^\prime\over Z_2}+7{Z_6^\prime\over Z_6} \right)\ . 
\end{equation}
Putting this all together gives the following pleasing results for the
stress--energy tensor at the discontinuity:
\begin{eqnarray}
S_{\mu\nu}&=&{1\over 2\kappa^2 \sqrt{G_{rr}}}
\left({Z_2^\prime\over Z_2}+{Z_6^\prime\over Z_6} \right) 
G_{\mu\nu}\ ,\nonumber\\
S_{ij}&=&0\ , \nonumber\\
S_{ab}&=&{1\over2\kappa^2\sqrt{G_{rr}}}
\left({Z_6^\prime\over Z_6} \right) 
 G_{ab} \ .
\label{stressed} 
\end{eqnarray}

Let us pause to admire the result.  The last line, referring to the
stress along the K3 direction, involves only the harmonic function for
the pure D6--brane part. This is appropriate, since there are only
D6--branes wrapped there. According to the middle line, there is no
stress in the directions transverse to the branes. This is consistent
with the fact that the constituent branes are BPS, and so there are no
inter--brane forces needed to support the shell in the transverse
space. 

As a first check of this interpretation, we can expand the results in
eqn.~(\ref{stressed}) for large $r_{\rm i}$. Up to an overall sign,
the coefficient of the metric components gives an effective tension
in the various directions. The leading contributions are simply:
\begin{eqnarray}
\tau_{\rm mem}(r_{\rm i})&=& {1\over 2\kappa^2}{r_6\over r_{\rm
i}^2}\left(1-{V_*\over V}\right) \nonumber\\
&=&{N\over (2\pi)^6
(\alpha^\prime)^{7/2}g_s}(V-V_*){1\over 4\pi r^2_{\rm
i}V}=N(\tau_6V-\tau_2)\left({1\over 4\pi r^2_{\rm i} V}\right)\
,\label{admcold}\\
\tau_{\rm K3}(r_{\rm i})&=& {1\over2\kappa^2}{r_6\over
r_{\rm i}^2}={N\over (2\pi)^6 (\alpha^\prime)^{7/2}g_s}\left({1\over
4\pi r^2_{\rm i}}\right)=N\tau_6\left({1\over 4\pi r^2_{\rm i}}\right)\ ,
\end{eqnarray}
which is in precise accord with expectations. In the K3 directions,
the effective tension matches precisely that of $N$ fundamental D6--branes,
with an additional averaging factor ($1/4\pi r^2_{\rm i}$) coming from
smearing the branes over the shell in the transverse space.
In the $x^{0,1,2}$ directions, we have an effective membrane tension which,
up to the appropriate smearing factor, again matches that for $N$ D6--branes
including the subtraction of $N$ units of D2--brane tension
as a result of wrapping on the K3 manifold\cite{bbg}.

We will make the matching of our effective stress tensor (\ref{stressed})
to a shell of D--brane sources more precise in the following section.
At this point, however, notice that the result for the
stress--energy in the unwrapped part of the brane is proportional to
$(Z_2Z_6)^\prime$. As we have already observed in 
(\ref{crucial}), this vanishes at precisely $r=r_{\rm e}$, and therefore we
recover the result\cite{jpp} that for incision at the enhan\c con
radius, there is a shell of branes of zero tension. 

For $r<r_{\rm e}$ we would get a negative tension from the
stress--energy tensor, which is problematic even in
supergravity. Notice however that nothing in our computation shows
that we cannot make an incision at any radius of our choosing for
$r\geq r_{\rm e}$, and place a shell of branes of the appropriate
tension (as in the calculation of the effective tensions at large $r_{\rm i}$
above).  This corresponds physically to the fact that probe branes
experience no potential, so they can consistently be placed at any
arbitrary position outside the enhan\c con. The special feature of the
enhan\c con radius in both cases is that it is a limit to where we can
place the branes. In section \ref{scatter}, we show that the enhan\c
con radius is also special from the point of view of particle scattering.

\subsection{Matching fields and branes}

We have seen that the effective tensions of the shell agree
with those expected for a collection of wrapped D6--branes at large radius,
and that the membrane tension vanishes at the same place as the
tension of the source branes. In fact, it is easily seen that the
stress--energy of the shell is in general precisely the same as the
stress--energy of $N$ wrapped D6--branes distributed uniformly on the
sphere, as we will now show\footnote{We
thank Neil Constable for useful discussions about these
matching calculations.}.

The integrated Einstein equation tells us that shell stress--energy
should be given by
\begin{equation}
S_{AB} = \int \sqrt{G_{rr}}\, dr\ \left[-
{{2}\over{\sqrt{-G}}} \sum_{\rm shell}
{{\delta S_{\rm{brane}}}\over{\delta G^{AB}}}\right]  \,,
\end{equation}
where the sum means that we should sum over the contributions of all
of the constituent branes in the shell.  The term in the square
brackets is just the standard definition for the stress--energy
tensor. As the source coming from the shell of branes is a
distribution (in the technical sense), the integral is included to
eliminate the radial $\delta$--function. Note that it is important
here that the variation is made with respect to the Einstein frame
metric.

The metric only appears in the Dirac--Born--Infeld part of the
D--brane action as can be seen in eqn.~(\ref{probeaction}).  Converting
this result to Einstein frame, the action for $N$ wrapped D6--branes
is
\begin{equation} \label{efact}
S_{\rm DBI} = - N\int_{{\cal M}_2}d^3\xi\, e^{-\Phi/4} (\mu_6 e^{\Phi} V_{\rm E}(r)
- \mu_2) (-\det{G_{\mu\nu}})^{1/2}, 
\end{equation}
where $V_{\rm E}(r)$ is the volume of the K3 in Einstein frame, {\it i.e.},
$V_{\rm E}(r) = \int_{K3} d^4 x \sqrt{-G_{ab}}$, and now ${G}_{\mu\nu}$ denotes
the pull--back of the Einstein frame metric to the effective membrane
worldvolume. We  assume that the
$N$ D6--branes are distributed uniformly over the $S^2$. Recall
that we  work in static gauge with $\xi^\mu=x^\mu, \mu =0,1,2$.
The stress--energy of these source branes can then be written as
\begin{eqnarray}
S_{\mu\nu} &=& {N e^{-\Phi/4} \over V_{\rm E}
{\rm Vol}(S^2)} (\mu_2-\mu_6
e^{\Phi} V_{\rm E} ) G_{\mu\nu} = {1 \over 2\kappa^2
\sqrt{G_{rr}}} \left( {Z_2' \over Z_2} + {Z_6' \over Z_6} \right)G_{\mu\nu}, 
\nonumber\\ 
S_{ab} &=& -{N e^{3\Phi/4} \over
{\rm Vol}(S^2)} \mu_6  G_{ab}
= {1 \over 2\kappa^2
\sqrt{G_{rr}}} {Z_6' \over Z_6} G_{ab}.
\label{trash}
\end{eqnarray}
Thus, we see that the form agrees with the shell stress--energy given
in eqn.~(\ref{stressed}).

Note that we have been slightly cavalier in doing the stress--energy
calculation using the effective membrane action in eqn.~(\ref{efact}).
The correct microscopic action would actually be that for the seven
dimensional worldvolume of the wrapped D6--branes. The term
proportional to $\mu_6$ implicitly takes this form since $V_{\rm E}$ is
defined as a four dimensional integral over the K3 surface.  However,
the contribution proportional to $\mu_2$ actually has its origin in an
$\alpha'$ correction to the standard DBI action for the
D6--branes\cite{bbg}. That is, implicitly this term involves an
integral over the K3 of an curvature--squared term and further this
combination of curvatures is not a topological invariant. Therefore
na\" \i vely it would appear that $S_{ab}$ should actually have
contributions proportional to $\mu_2$. In fact, however, this
contribution vanishes because~K3 is Ricci--flat. The difference
between the curvature--squared interaction in the D6--brane action and
the Gauss--Bonnet invariant is proportional to $2 R_{ab}R^{ab}-R^2$.
Hence the nontrivial contributions coming from the variation of these
terms will be proportional to the Ricci tensor or Ricci scalar, and so
vanish when evaluated on K3.  Similarly the contributions proportional
to $\mu_2$ in the effective membrane action (\ref{probeaction})
originate from the integral over K3 of a curvature--squared term in
the Wess--Zumino part of the D6--brane action. In this case, these
anomaly induced terms do form a four dimensional topological
invariant, the first Pontryagin class\cite{anom}.  Hence there is no
contribution to the stress tensor coming from the variation of these
terms either.

Hence these results (\ref{trash}) provide a further verification that
matching the repulson solution (\ref{outside}) to a flat space
interior (\ref{inside}) at any radius has the interpretation of
introducing a shell of wrapped D6--branes as the source. As a
additional check, we can also consider the matching of the other
fields. The simplest to consider are the R--R fields.  Since the
exterior geometry contains $N$ units of 7--form flux and $-N$ units of
3--form flux, and the interior has none, the shell clearly carries the
same R--R charges as $N$ wrapped D6--branes.

It is interesting and instructive to consider the junction conditions
for the dilaton in detail. Since this issue is rarely discussed, we
begin with the simple case of a shell of extremal D6--branes living
unwrapped in flat ten dimensional spacetime. The dilaton and R--R
fields are written in terms of the harmonic function $\tZ_6$ as
\begin{equation}
{{e^{\Phi}}} = {g_s}\tZ_6^{-3/4} \,,\quad
C_{(7)} = ({g_s\tZ_6})^{-1} dx^0\wedge dx^1\wedge dx^2
\wedge V\,\varepsilon_{\rm K3} \,,
\end{equation}
while the metric is, in Einstein frame,
\begin{equation}
ds^2=\tZ_6^{-1/8}\left(-dt^2+\sum_{i=1}^6 dx_i^2\right)+
\tZ_6^{7/8} (dr^2+r^2d\Omega_2^2) \ .
\end{equation}
In this BPS case, the radial component of the metric is continuous at
the shell, but this is not true generally.  The only requirement is
that the induced metric transverse to $r$ be continuous at the shell.

Since the function $\tZ_6$ is harmonic, for a shell of D6--branes
at radius $r_{\rm i}$, Gauss's law demands
\begin{equation}
\tZ_6(r)=Z_6(r_{\rm i}) + \theta(r-r_{\rm i}) \left[ Z_6(r)-Z_6(r_{\rm i})
\right] \ ,
\end{equation}
where $Z_6(r)$ is as given in eqn.~\reef{harms}.
Hence, differentiating with respect to $r$,
\begin{equation}
\tZ_6'(r)= \theta(r-r_{\rm i}) Z_6'(r)
\end{equation}
and once again
\begin{equation}
\tZ''_6(r)=\theta(r-r_{\rm i}) Z''_6(r) + \delta(r-r_{\rm i}) Z'_6(r) \,.
\end{equation}
It is the singular delta function which gives rise to the junction
conditions.  We therefore need to find all the places in which
$\tZ''_6$ appears.

For the bulk dilaton equation of motion we have, given the  electric
coupling to the R--R  potential $C_{(7)}$
\begin{equation}
\nabla^2 \Phi - {{(-3/2)}\over{2(8)!}}e^{-3\Phi/2}
(\partial C)^2 =-{{2\kappa^2}\over{\sqrt{-G}}}
{{\delta}\over{\delta\Phi}} \sum_{\rm shell} S_{\rm brane} \,.
\end{equation}
For the purpose of discovering junction conditions, our only bulk concern
here is the Laplacian of the dilaton; the R--R field strength term has
too few derivatives.  For the left hand side of the bulk equation we
therefore have for the singular term just
\begin{equation}
G^{rr}\Phi''\simeq -{{3}\over{4}}G^{rr}\tZ_6(r)^{-1}\tZ''_6(r) \simeq-{{3}\over{4}}
{{1}\over{G_{rr}}}\tZ_6(r)^{-1} \ 
\delta(r-r_{\rm i}) \tZ'_6(r)  \,,
\end{equation}
or in covariant language
\begin{equation}
\nabla^2\Phi \simeq-{{3}\over{4}}
 {\tZ_6(r)}^{-1} \  
n^r \nabla_r \tZ_6(r)\ 
{{\delta(r-r_{\rm i})}\over{\sqrt{G_{rr}}}} \simeq
 n^r\nabla_r \Phi {{\delta(r-r_{\rm i})}\over{\sqrt{G_{rr}}}} \,.
\end{equation}
More generally, we can encode the covariant integrated discontinuity as
\begin{equation} \label{dil1}
2\kappa^2 S_\Phi \equiv (n^r\nabla_r \Phi|_{r=r_{\rm i} + \epsilon}
-n^r\nabla_r \Phi|_{r=r_{\rm i} - \epsilon}  )
= {{1}\over{\sqrt{G_{rr}}}} (\Phi'|_{r=r_{\rm i} + \epsilon} -
\Phi'|_{r=r_{\rm i} - \epsilon}) \,.
\end{equation}
For the brane source term, we begin with the usual DBI
brane action, as the Wess--Zumino term couples only to the bulk R--R
field.  The D6--branes of the shell are distributed evenly over the
transverse two--sphere ($\Omega$), so that
\begin{equation}
\begin{array}{l}\bs
{\displaystyle{
-{{2\kappa^2}\over{\sqrt{-G}}} 
{{\delta}\over{\delta\Phi(y)}} \sum_{\rm shell}
S_{\rm brane}  }} \cr\bs
=  {\displaystyle{
2\kappa^2
\left( \int d^2\Omega \,
{{N}\over{4\pi}} \right) \left[
{{\mu_6}\over{g_s}} \int d^7\xi
\left({{3}\over{4}} \right) e^{3\Phi(x)/4} 
\sqrt{-\det{G_{\mu\nu}}(x)} \right] 
{{\delta^{10}(y-x(\xi))}\over{{\sqrt{-G}}}} }} \cr
= {\displaystyle{
{{3\kappa^2\mu_6N}\over{8\pi g_s}}\,
{{\delta(r-r_{\rm i})}\over{\sqrt{G_{rr}G_{\Omega\Omega}}}} \,
e^{3\Phi(r)/4} }} \,.
\end{array}
\end{equation}
Here it is important that the variation of the dilaton is made
while holding the Einstein frame metric fixed.
The BPS D6--brane shell therefore gives the source
\begin{equation} \label{dil2}
2\kappa^2 S_\Phi=
{{3\kappa^2\mu_6N}\over{8\pi g_s}}\,
{{e^{3\Phi(r)/4}}\over{\sqrt{G_{\Omega\Omega}}}}  \,.
\end{equation}
We want to show that eqns. (\ref{dil1}) and (\ref{dil2}) agree. Using $\Phi'
= -3 H_6'/4 H_6$, and continuity of $H_6$, this will be true if 
\begin{equation} \label{dilcond}
-(\tZ'_6(r)|_{r=r_{\rm i} + \epsilon} - \tZ'_6(r)|_{r=r_{\rm i} - \epsilon})  
= {{\kappa^2\mu_6N}\over{2\pi g_s}}  \ 
{{{\tZ_6(r)\sqrt{G_{rr}}}
e^{3\Phi(r)/4}}\over{\sqrt{G_{\Omega\Omega}}}}  
= 2\kappa^2{{N\mu_6}\over{4\pi r_{\rm i}^2}g_s}. 
\end{equation}
Note that the factors of $\tZ_6$ have cancelled in a non--trivial
manner. Thus, this result is consistent with the usual form of the
harmonic function; \ie eqn. (\ref{dilcond}) is satisfied if $Z_6$
takes its usual form,
\begin{equation}
\mu_6=(2{\pi})^{-6}{\alpha^\prime}^{-7/2}\longrightarrow
Z_6(r)=1+{{r_6}\over{r}}\ ,\qquad r_6 \equiv { g_sN{\alpha^\prime}^{1/2}
 \over2} \ .
\end{equation}
Thus, for a spherical shell of unwrapped D6--branes, the discontinuity
in derivative of the dilaton field at the shell agrees with the source
term in the brane worldvolume action.

To extend this to the case of
the enhan\c con is straightforward. The crucial point again is that
we need to consider the DBI action in
the Einstein frame. Using eqn.~(\ref{efact}), the wrapped D6--brane shell
then gives the source
\begin{equation}
2\kappa^2 S_{\Phi} = 
{{\kappa^2 N}\over{2\pi g_s}}\,
{1\over{\sqrt{G_{\Omega\Omega}}}} \left( {3 \over 4}
{e^{3\Phi(r)/4}}\mu_6 + {1 \over 4} 
{e^{-\Phi(r)/4}\over V_{\rm E}(r)}\mu_2 \right)
\,.
\end{equation}
Note that while this result coincides with that expected from the
effective membrane action (\ref{efact}), it can also be properly
derived with the curvature--squared interactions appearing in the
D6--brane action\cite{bbg,anom}, assuming that the curvatures are
calculated with the string frame metric.  The dilaton discontinuity in
this case is given by
\begin{equation}
2\kappa^2 S_\Phi = {{1}\over{\sqrt{G_{rr}}}} (\Phi'|_{r=r_{\rm i} + \epsilon} -
\Phi'|_{r=r_{\rm i} - \epsilon}) = {1 \over \sqrt{G_{rr}}} \left( -{3
\over 4} {Z_6' \over Z_6} + {1 \over 4} {Z_2' \over Z_2} \right), 
\end{equation}
where $Z_6, Z_2$ are given in eqn.~(\ref{harms}). Thus, we see that
behaviour of the dilaton field at $r=r_{\rm i}$ is correctly accounted
for by a source shell of wrapped D6--branes. One particular point to
note is that the shell still acts as a source for the dilaton at the
enhan\c con radius where the effective membrane tension vanishes.

\subsection{Particle Scattering}
\label{scatter}

There is another reason why the enhan\c con radius is special in
supergravity. It is the place where the string frame metric begins to
show repulsive behaviour for geodesic probes. That is, we probe the
solution with particles which feel only the (string frame) geometry
and do not have any additional couplings to the dilaton or R--R fields.
This is a natural probe computation to perform when considering
massive string modes.

There is a pair of Killing vectors, \boldmath${\xi}$\unboldmath
$\,\,=\partial_t$ and \boldmath${\eta}$\unboldmath
$\,\,=\partial_\phi$, and so a probe with ten--velocity $\mathbf{u}$
has conserved quantities $e=-$\boldmath${\xi}\cdot{u}$\unboldmath $\,$
and $\ell=-$\boldmath$\eta\cdot{u}$\unboldmath $\,$. $e$ and $\ell$
are the total energy and angular momentum per unit mass, respectively.
We have frozen the motion in the longitudinal directions.

The radial evolution is given by one dimensional motion in an
effective potential with 
\begin{eqnarray} 
{dr\over d\tau}&=&\pm\sqrt{E-V_{\rm eff}(r)}\ ,\nonumber\\ {\rm
where}\quad V_{\rm eff}(r)&=&{1\over2}\left[{1\over
G_{rr}}\left(1+{\ell^2\over G_{\phi\phi}}\right)-1\right]\ ,\qquad
E={e^2-1\over 2}\ ,
\label{radial}
\end{eqnarray}
where the metric components in the above are in string frame.  Let us
specialise to the study of purely radial motion, with zero impact
parameter. We can see if the geometry will repel the probe at some
radius by simply placing it there at rest, and observing if it rolls
towards larger or smaller $r$. For large enough $r$, the effective
potential is indeed attractive, and so we need only seek a vanishing
first derivative of $V_{\rm eff}(r)$.  This gives the condition:
\begin{equation}
G_{tt}^\prime=0\ ,
\end{equation}
which is in fact condition (\ref{crucial}), and so we see that the
particle begins to be repelled precisely at $r=r_{\rm e}$. Particles
with non--zero angular momentum will of course experience additional
centrifugal repulsion, but $r=r_{\rm e}$ is the boundary of the region
where there is an intrinsic repulsion in the geometry.  See the top
curve in figure~\ref{veff}.

Cutting the geometry at any smaller radius would leave a region where
the geometry is repulsive and so we see that the enhan\c con radius is
therefore the minimal radius at which we can excise all of the
infection inherited from the repulson\footnote{This question,
  essentially ``Just how much repulsion is left over after excision?''
  was raised by audience members (Scott Thomas, Lenny Susskind, Matt
  Kleban, John McGreevy, and possibly others) during a lecture on
  enhan\c cons by CVJ at the ITP Stanford. We thank them for the
  question.}.

A similar computation can be done for the massless modes too, with the
following result for the effective motion:
\begin{eqnarray}
{e^2\over\ell^2}&=& {1\over\ell^2}\left({dr\over
d\lambda}\right)^2+Q_{\rm eff}(r)\ ,\nonumber\\ {\rm where}\qquad
Q_{\rm eff}(r)&=&{1\over {Z}_2(r){ Z}_6(r) r^2}\ .
\end{eqnarray}
The effective potential $Q_{\rm eff}(r)$ is purely centrifugal, and
represents the usual bending of light rays by the geometry.

So we conclude that the part of extremal geometry which truly should
be called the ``repulson'' geometry actually begins at $r=r_{\rm e}$.
Hence in establishing the enhan\c con at precisely this radius, string
theory avoids creating in these configurations a region of the
spacetime which is both naked (\ie not surrounded by an event horizon)
and intrinsically repulsive.

\section{Adding D2--Branes}
\label{d2}

A useful generalisation\footnote{The analogous construction for the
  D5/D1--brane system wrapped on K3 was considered in
  ref.~\cite{robme} in considering the role of the enhan\c con
in the physics of
extremal black holes.}  is to add D2--branes to the enhan\c con
configuration described above\footnote{We remind the reader that the
  harmonic functions $Z_2$ in the previous geometry are nothing to do
  with real D2--branes, but arise as a result of the induced D2--brane
  charge produced by wrapping on the K3.}. They preserve the same
amount of supersymmetry as the original configuration, and so it is
easy to compute how a single D2--brane probe sees the enhan\c con
geometry.  Since a D2--brane probe does not have any sensitivity to
the K3 part of the geometry, there is no enhan\c con effect for it, and it
can travel all the way in to the origin at $r=0$ \cite{fuzzy}. Hence
we can imagine building up D2--branes inside the enhan\c con radius.
We will also find that the presence of the D2--branes actually allows
a certain fraction of the D6--branes to move inside the enhan\c con
shell. Therefore we present a solution below describing a system of $N$
wrapped D6--branes and $M$ D2--branes. Of these, $N'$ of the
D6--branes and all $M$ of the D2--branes are placed at $r=0$ while
$N-N'$ of the D6--branes remain in the enhan\c con shell. After
presenting the solution, we will discuss the physics of these
configurations.

\subsection{The Geometry}

Given a spherically symmetric configuration of $N$ wrapped D6--branes,
the effect of adding $M$ real D2--branes which are smeared over the K3
is to increase the D2--brane charge from $-N$ to $M-N$. In the
exterior solution, this shift is simply accomplished by modifying the
harmonic function $Z_2$ in eqn.~(\ref{harms}). The scale appearing
there is now 
\begin{equation} r_2=r_6{V_*\over V}\left(1-{M\over
      N}\right)\ ,
\label{newarr}
\end{equation}
while the scale $r_6$ remains as in eqn.~(\ref{arrs}).
The enhan\c con radius is now given by the slightly more general expression:
\begin{equation}
r_{\rm e}={2V_*r_6\over V - V_*}\left(1-{M\over2N}\right)\ ,
\label{newR}
\end{equation}
and the exterior solution with the modified $r_2$ applies for $r>
r_{\rm e}$.  Notice that eqn.~(\ref{newR}) seems to indicate that the
enhan\c con shell shrinks due to the presence to the D2--branes. One
can easily verify that as well as the coordinate position of the shell
becoming smaller, the proper area of the shell also becomes smaller.
In particular for $M\ge 2N$, there is no enhan\c con shell at all.
That is, the D6--branes and D2--branes can all coalesce to a
point--like configuration at the origin.

As in the previous constructions, we will describe the
incision at  some arbitrary radius $r_{\rm i}\ge r_{\rm e}$.
We match onto an interior solution with $N'$ D6--branes and $M$
D2--branes placed at the origin. Thus the interior geometry is
given by
\begin{equation}
g_s^{1/2}\,ds^2 = H_2^{-5/8} H_6^{-1/8} \eta_{\mu\nu} dx^\mu dx^\nu +
H_2^{3/8} H_6^{7/8} (dr^2 + r^2 d\Omega) + V^{1/2}
H_2^{3/8} H_6^{-1/8} ds^2_{K3}
\end{equation}
and the non--trivial fields are
\begin{eqnarray}
e^{2\Phi} &=& g_s^2 H_2^{1/2} H_6^{-3/2}\ , \nonumber\\
C_{(3)} &=& (g_s H_2)^{-1} dx^0
\wedge dx^1 \wedge dx^2\ ,\nonumber\\ 
C_{(7)} &=& (g_s H_6)^{-1} dx^0\wedge dx^1 \wedge dx^2
\wedge V\,\varepsilon_{\rm K3}\ ,
\end{eqnarray}
where
\begin{eqnarray}
H_2 = 1-{r_2 - r_2'\over r_{\rm i}} - {r_2' \over r}, 
\qquad r_2^\prime= r_6{V_*\over V}{N'-M\over N}\ ,
 \label{h2}\\
H_6 = 1+{r_6 - r_6'\over r_{\rm i}} + {r_6' \over r}, 
\qquad r_6^\prime= r_6{N'\over N}={g_sN'\alpha^{\prime1/2}\over2}\ ,
 \label{h6}
\end{eqnarray}
where the constant terms in the harmonic functions are chosen to ensure
continuity of the solution at the incision radius, $r=r_{\rm i}$. This interior
solution is, of course, essentially the same as the exterior solution
with modified harmonic functions.

After computations analogous to those of the previous section, we get a
stress tensor
\begin{eqnarray}
2 \kappa^2 S_{\mu\nu} &=&  {1 \over \sqrt{G_{rr}}} \left( {Z_2' \over
Z_2} + {Z_6' \over Z_6} - {H_2' \over H_2}- {H_6' \over H_6} \right)
G_{\mu\nu} \ , \nonumber\\
S_{ij}  &=&0\ , \nonumber\\
2 \kappa^2 S_{ab} &=&   {1 \over \sqrt{G_{rr}}} \left({Z_6' \over Z_6}
- {H_6' \over H_6}\right) G_{ab}.
\label{newstresses}
\end{eqnarray}
We see once again that the pressure in the shell directions vanishes, in
agreement with the fact that this system is still BPS. Furthermore, we can
show that the effective tension in the $x^{0,1,2}$ directions vanishes
precisely at the enhan\c con radius, and more generally the discontinuities 
at the shell agree with the source terms in the worldvolume action
of $N-N'$ wrapped D6--branes. 

\subsection{The Physics}

Let us consider how the configuration above could be constructed
physically. For the following physics discussion, we will only consider the
case where the shell sits at precisely $r_{\rm i}=r_{\rm e}$, with
the enhan\c con radius given in eqn.~(\ref{newR}).  We will further
assume that $M<2N$ in order that there is an  enhan\c con.

First beginning with only wrapped D6--branes in an enhan\c con shell,
we noted above that a D2--brane encounters no obstacle to moving past
$r=r_{\rm e}$ to the origin. Hence a natural solution to think about
is that with $M$ D2--branes at the origin and all $N$ wrapped
D6--branes in the shell, \ie $N'=0$. In this case, $r_6^\prime=0$ and
so the six--brane harmonic function is simply a constant,
$H_6=Z_6(r_{\rm e})$. As defined in eqn.~(\ref{h2}), $r_2$ is negative
for this configuration, and so the two--brane harmonic function takes
the form $H_2=\gamma+|r_2'|/r$, which grows as $r$ decreases. Now the
volume of the K3 is given by
\begin{equation}
V(r)=V{H_2\over H_6}=V{\gamma+{|r_2'|/ r}\over Z_6(r_{\rm e})}
\label{trashv}
\end{equation}
where the constants in the expression are fixed by the condition
$V(r_{\rm e})=V_*$.  Now the interesting observation is that $V(r)$ is
a function that grows as $r$ decreases in the interior solution.
Therefore there is no obstruction to some of the wrapped D6--branes
migrating from the enhan\c con shell to the origin.

Examining the solution above further, one finds that the growth of the
K3 volume noted above is suppressed when we begin to place
D6--branes at the origin. Hence, we might ask at what point this growth
stops altogether. As $V(r)$ is a monotonic function, the answer is
most easily determined by requiring:
\begin{equation}
V(r=0)=V{H_2(0)\over H_6(0)}=V{|r_2'|\over r'_6}=V_*
\label{trashvv}
\end{equation}
from which we find $N'=M/2$. With this choice of parameters, we have
in fact that $H_2(r)= V_*/V\,H_6(r)$ which produces some
simplifications in the solution above. The most important point,
however, is that as measured in the string frame, the volume of the K3
is a fixed constant $V_*$ for the interior solution.  Hence, given this
configuration with $N'=M/2$, a probe D6--brane can not move into the
interior, and so any D6--branes which might be added to this
configuration would accumulate at the enhan\c con shell.

Thus one might think that this is a limiting configuration. However,
it is still possible to move {\it additional} D6--branes inside the
enhan\c con\footnote{We thank Joe Polchinski for a very useful
  discussion of this point.}.  In principle, this is achieved by
carrying some of the D2--branes back out to the enhan\c con shell, and
by binding them to an equal number of the D6--branes there. This
composite unit can now move below $r_{\rm e}$, since the negative
tension induced from wrapping the K3 is cancelled by the tension of
the instantonic D2--branes smeared over the worldvolume of the
D6--branes. This threshold bound state becomes a true BPS bound state
for $r<r_{\rm e}$, since separating them would produce a pure
D6--brane which is not BPS in that region --- see ref.~\cite{robme}
for further discussion. Through this mechanism we can also form
configurations with up to $N'=M$ units of D6--brane charge located at
$r<r_{\rm e}$. Thus the true limiting solution is that with $N'=M$.
That is, our reasoning using string theory facts would indicate that
the supergravity solutions with $N'>M$ are unphysical.

The latter conclusion is also supported by the fact that in this case
$r'_2>0$, and so there is a repulson singularity in the geometry.  The
repulsive nature of the singularity is seen from the following
supergravity calculation: Consider an excision construction where flat
space is inserted as the interior solution, \ie the shell contains N
D6--branes and M D2--branes with $N>M$, and calculate the associated
shell stress--energy, \ie set $H'_{2,6}=0$ in eqn.~\reef{newstresses}.
One would find that before the shell could shrink down to zero size,
the tension would become negative indicating an unphysical
configuration was reached. Hence even supergravity alone seems to
regard as unphysical the repulson--like configurations with $N$ or
$N'>M$.
 
In the limiting case, the interior solution has no net D2--brane
charge, as can be seen from the fact that $r'_2=0$.  Furthermore, one
finds that the K3 volume actually does shrink to below the stringy
value, $V_*$, for this configuration (or any solution with $M/2\le
N'<M$).

Perhaps a few final remarks are in order here: First, the enhan\c con
radius (\ref{newR}) remains unaffected by the migration of D6--branes
from the enhan\c con shell to the origin.  The second comment is that
we are considering a BPS system of branes, and even within the
restriction to spherical symmetry, we could generalise these solutions
to having concentric shells of combinations of D6-- and D2--branes
both inside and outside the enhan\c con.

\section{Some Non--Extremal Enhan\c cons}
\label{gooddef}

Having learned about the nature of the enhan\c con and the excision
process in the extremal case, where we have the aid of supersymmetry
to guide our intuition, we now feel able to proceed towards the
unknown, and study the non--extremal geometry. What we have in mind is
applying some of our new--found experience to studying the physics of
the enhan\c con at finite temperature. Some suggestions with respect
to this topic were made in ref.~\cite{jpp}. We generalise the
discussion by adding D2--branes to the system of wrapped D6--branes.
If the number of D2--branes $M$ exceeds the number of D6--branes $N$,
the net D2--charge is positive and the non--extremal solutions
familiar from the unwrapped case can be adapted to the present
problem. Hence we begin by studying these solutions in the following
two sub--sections. In fact, these solutions also accommodate the
situation where $M<N$, as we will show in sub--section~\ref{funnydef}.
Taking $M$ to zero recovers the proposed solution for the
non--extremal enhan\c con considered in ref.~\cite{jpp} (up to the
correction of a small typographical error).  However, upon closer
examination, we find that these solutions exhibit certain
peculiarities which makes their physical interpretation uncertain, as
discussed in sub--section~\ref{funnyfizz}.

\subsection{The Geometry}

We study the non--extremal solution describing $N$ wrapped D6--branes
and $M$ D2--branes. As we said above, we limit ourselves here to the
case $M\ge N$ so that the net D2--brane charge is positive (or zero). Then
the exterior solution takes the familiar non--extremal form. 
So for $r>r_{\rm i}$, we have (in Einstein frame):
\begin{eqnarray}
g_s^{1/2}\,ds^2  &=&\Z_2^{-5/8}{\Z}_6^{-1/8} (-K dt^2 + dx_1^2+ dx_2^2) +
\Z_2^{3/8}\Z_6^{7/8} (K^{-1} dr^2 + r^2 d\Omega_2^2) \nonumber\\
&\phantom{+}&\hskip5cm + V^{1/2} \Z_2^{3/8}\Z_6^{-1/8}ds_{\rm K3}^2\ .
\label{junterior}
\end{eqnarray}
The dilaton and R--R fields are 
\begin{eqnarray}
e^{2\Phi} &=& g_s^2\Z_2^{1/2}\Z_6^{-3/2}\ , \nonumber \\
C_{(3)} &=& (g_s \alpha_2 \Z_2)^{-1}dt \wedge dx^1 \wedge dx^2\ ,
\nonumber\\
C_{(7)} &=& (g_s \alpha_6 \Z_6)^{-1} dt \wedge dx^1 \wedge dx^2 \wedge
V\,\varepsilon_{\rm K3}\ .
\end{eqnarray}
The various harmonic functions are given by 
\begin{eqnarray}
K &=& 1 - {r_0 \over r}\ ,\nonumber\\ 
\Z_2&=&1 + {\alpha_2 r_2 \over r}\ ,\qquad \alpha_2 = - {r_0 \over 2r_2} +
\sqrt{1 + \left( {r_0 \over 2r_2} \right)^2}\ ,
\nonumber\\  
\Z_6&=& 1 + {\alpha_6 r_6 \over r}\ ,\qquad \alpha_6 = - {r_0 \over 2r_6} +
\sqrt{1 + \left( {r_0 \over 2r_6} \right)^2}\ ,
\label{z6ext}
\end{eqnarray}
where 
\begin{equation}
r_2=r_6{V_*\over V}\left({M\over N}-1\right)\ ,
\label{juncale}
\end{equation}
and $r_6$ is still as given in eqn.~\reef{arrs}. Note that the present
definition of $r_2$ differs by a sign from that given in eqn.~\reef{newarr}
in the previous section, so that this quantity is positive throughout the
current discussion where
$M\ge N$. The enhan\c con radius is the place where the running volume of
K3 gets to the value $V_*$. The result may be written as
\begin{equation}
\re = {V_*\alpha_6 r_6 -V \alpha_2 r_2 \over V - V_*}\ .
\label{junkre}
\end{equation}
As we found before, the presence of the D2--branes causes the enhan\c
con radius to shrink, and so with a sufficiently large number of
D2--branes, there will be no enhan\c con shell outside the horizon at
$r=r_0$. Furthermore, note that if $r_0$ is large enough, $\re$ will
fall inside the horizon. In fact, $r_0 = \re$ at
\begin{eqnarray}
r_0^2 &=&{ (V_*^2r_6^2-V^2r_2^2)^2 \over V_*(V-V_*)(V_*r_6^2-r_2^2)}
\nonumber\\
&=&{ V_*^2r_6^2\left(2-(M/N)^2\right)^2(M/N)^2 \over (V-V_*)
(V-V_*(1-M/N)^2)}\ . \labell{bigjunk}
\end{eqnarray}
Note that this result only applies for $M<2N$. As we noted before for
$M=2N$, the enhan\c con radius collapses to zero with $r_0=0$. For
larger values $M\ge 2N$, the enhan\c con radius is always inside the
horizon (and the result in eqn.~\reef{bigjunk} is spurious).  For
small enough $r_0$ and $r_2$ (or $M/N$), we will have $\re > r_0$, and
we can have a solution with an enhan\c con shell. Given
eqn.~\reef{bigjunk}, it is useful to keep in mind then that we are
thinking of $r_0\simle V_*/V\,r_6$. By our procedures of the previous
section, we match on to some nontrivial interior geometry which is
composed of a non--extremal system of $M$ D2--branes and $N'$ wrapped
D6--branes.

Hence we take the solution inside some incision radius $r_{\rm i}$ to
 be of the form:
\begin{eqnarray}
g_s^{1/2}\,ds^2 &=& \hH_2^{-5/8} \hH^{-1/8} \left(- {K(r_{\rm i})\over
L(r_{\rm i})}L dt^2 + dx_1^2+dx_2^2\right) + \hH_2^{3/8} \hH_6^{7/8}
(L^{-1} dr^2 + r^2 d\Omega) \nonumber\\
&\phantom{+}&\hskip5cm + V^{1/2} \hH_2^{3/8} \hH_6^{-1/8}ds^2_{K3}\ ,
\label{tag1}
\end{eqnarray}
with accompanying fields
\begin{eqnarray}
e^{2\Phi} &=& g_s^2 \hH_2^{1/2} \hH_6^{-3/2}\ , \nonumber\\
 \qquad C_{(3)} &=&
\left({K(r_{\rm i})\over L(r_{\rm i})}\right)^{1/2} (g_s
 \alpha'_2\hH_2)^{-1} dt \wedge dx^1 \wedge dx^2\ , \nonumber\\
C_{(7)} &=& \left({K(r_{\rm i})\over L(r_{\rm i})}\right)^{1/2}
(g_s \alpha'_6 \hH_6)^{-1} dt\wedge dx^1 \wedge dx^2 
\wedge V\,\varepsilon_{\rm K3}\ ,
\end{eqnarray}
where
\begin{eqnarray}
L &=& 1 - {r_0' \over r}\ ,\nonumber\\
\hH_2 &=& 1 +{\alpha_2r_2-\alpha_2'r_2'\over r_{\rm i}} +
{\alpha_2' r_2' \over r}\ ,
\nonumber\\
&&\qquad\qquad\alpha_2' = - {r_0' \over 2r_2'} +\sqrt{1 + 
\left( {r_0' \over 2r_2'} \right)^2}\ ,
\qquad r_2'=r_6{V_*\over V}{M-N'\over N}\ ,
\nonumber\\
\hH_6 &=& 1 +{\alpha_6r_6-\alpha_6'r_6'\over r_{\rm i}} +
{\alpha_6' r_6' \over r}\ ,
\nonumber\\
&&\qquad\qquad\alpha_6' = - {r_0' \over 2r_6'} +\sqrt{1 + 
\left( {r_0' \over 2r_6'} \right)^2}\ ,
\qquad r_6'=r_6{N'\over N}\ .
\label{tag2}
\end{eqnarray}
Note that we have introduced an independent non--extremality
scale $r_0'$ for the interior solution. Implicitly $r_0'<r_{\rm i}$
in order that the interior black hole actually fits inside the
shell. Also note that we have been
lax in imposing continuity at the shell. Our choice of
constants in the harmonic functions ensures that the geometry and metric
are continuous there, but R--R potentials jump by a constant term. While
the latter is pure gauge, it means that in a probe calculation the probe
may acquire a spurious constant energy in crossing the shell --- however,
we will not present any such calculations here.

The stress tensor of the shell is now
\begin{eqnarray}
2 \kappa^2 S_{tt} &=&  {1 \over \sqrt{G_{rr}}} \left[{\Z_2' \over \Z_2}
+{\Z_6' \over \Z_6} +{4 \over r_{\rm i}}-
\sqrt{L \over K}
\left({\hH_2'\over \hH_2}+{\hH_6'\over \hH_6}+{4 \over r_{\rm i}}\right)
\right]G_{tt}\ ,\nonumber\\
2 \kappa^2 S_{\mu\nu} &=&  {1 \over \sqrt{G_{rr}}} \left[{\Z_2' \over \Z_2}+
{\Z_6' \over \Z_6} + {K' \over K} + {4 \over r_{\rm i}}
 -\sqrt{L \over K}\left({\hH_2'\over \hH_2}+{\hH_6'\over
\hH_6}+ {L' \over L}+ {4 \over r_{\rm i}}\right)
  \right] G_{\mu\nu}\ ,\nonumber\\
2 \kappa^2 S_{ij}  &=&  {1 \over \sqrt{G_{rr}}} \left[{K' \over K}+ {2 \over
r_{\rm i}} - \sqrt{L \over K}\left({L' \over L} + {2 \over r_{\rm i}}\right)
 \right] G_{ij}\ ,\label{trashij}\\ 
2 \kappa^2 S_{ab} &=&   {1 \over \sqrt{G_{rr}}} \left[{\Z_6' \over \Z_6} +
{K' \over K} + {4 \over r_{\rm i}}- \sqrt{L \over K}
\left({\hH_6'\over \hH_6}+{L' \over L}+{4 \over r_{\rm i}}\right)\right]
G_{ab}\ .
\nonumber
\end{eqnarray}
Here, the indices on $S_{\mu\nu}$ only run over the $x^1$ and $x^2$
directions. Furthermore, in these expressions, $G_{rr}$ denotes the radial
component of the exterior metric as given in eqn.~\reef{junterior}.

\subsection{Some Physics}

For the above solution, we will regard $r_0,$ $r_2$ and $r_6$ (or
alternatively $r_0,$ $M$ and $N$) as fixed parameters as they define
the mass and R--R charges of the given configuration. This leaves
$r_0'$ and $N'$, as well as the incision radius $r_{\rm i}$,
as free parameters, which we expect should be fixed by the physics
of the enhan\c con. Also recall that we are working with $N\le M<2N$.
The first bound is required for the asymptotic D2--charge to be positive.
The second bound is imposed in order that the enhan\c con radius can
appear outside of the event horizon at $r=r_0$.

To gain some intuition for these solutions, we imagine that they
arise by beginning with a (spherically symmetric) BPS configuration
of D2-- and D6--branes, and then introducing
a small Schwarzschild black hole at the centre. The black hole
upsets the balance of forces of the original configuration, and so
the branes begin to fall towards the origin. Following the discussion
of section 3, there is no mechanism to halt the infall of the
D2--branes and so we have implicitly assumed that they are all
sucked into the black hole --- that is, we have set the number
of D2--branes in the interior solution to $M$, the total number
of D2's in the system. It is straightforward to show that in the
non--extremal background the tension of a probe D6--brane still becomes
negative precisely when the K3 volume shrinks below $V_*$. Hence
although there seems to be no obstacle for the wrapped D6--branes to
reach $r=\re$, the enhan\c con provides a potential mechanism
to restrain the further infall of these branes. Hence for our solution
to be physically relevant, it seems we must fix the incision radius
to match the enhan\c con radius \reef{junkre}. Notice that the latter
is completely fixed by the exterior parameters: $r_0,$ $M,$ $N$.

Therefore, one might be tempted to think about a solution with $N'=0$,
\ie the solution where all of the D6--branes are fixed in the shell at
$r=\re$. However, just as in section 3.1, we consider the volume of
the K3 in the interior region: \beq \widehat{V}(r)=V{\hH_2\over\hH_6}
\label{junkvol}
\eeq where with $r_{\rm i}=\re$, we have ensured that
$\widehat{V}(\re)=V_*$.  With the choice $N'=0$, we note that
$\hH_6$ is a fixed constant while $\hH_2$ grows with decreasing
radius. Hence once again, it seems that there is no obstacle for some
of the D6--branes to fall from the shell into the black hole at the
centre. This process would have to continue at least until the point
where $V(r)$ is constant over the entire interior region. This
condition is most easily determined by setting $\widehat{V}(r=0)=V_*$
which yields \beq {\alpha_2'r_2'\over\alpha_6'r_6'}={V_*\over V}
\label{consvol}
\eeq which implicitly gives a constraint relating $r_0'$ and $N'$.
While it is a straightforward algebraic exercise to explicitly
determine, $N'=N'_*(r_0')$, the result is not particularly
illuminating.  However, we note that for small $r_0'$ the approximate
result takes the form \beq N'\simeq{M\over
  2}\left(1-{1\over4}\left(1-{V_*\over V}\right){V r_0'\over V_*
    r_6}+\cdots\right)\ .
\label{junkstuf}
\eeq This result indicates that introducing the black hole actually
makes the interior region less accessible to the wrapped D6--branes.
That is, in the limit \reef{consvol}, the critical number of
D6--branes is less in the non--extremal case than in the BPS
configuration, $N'=N'_*<{M/ 2}$.

Fixing our last free parameter $r^\prime_0$ seems to require more
subtle physical insight. One approach to further fix $r_0'$ is to
consider the components of the shell stress--energy \reef{trashij} in
the transverse space. In particular, the sign of these components
is completely determined by $r_0$ and $r_0'$.  For $r_0'>r_0$, the sign
is negative and so there is a positive tension in these directions.
That is, the shell appears to want to expand in the transverse space,
but it is held in place by an internal tension.  For $r_0'<r_0$, the
sign is positive and so there is a positive pressure in these
directions. Hence in this case, the shell appears to want to collapse
in the transverse space, but is held in place by an internal pressure.
Given the physical intuition that the imbalance of forces is caused by
the gravitational attraction of the central black hole, it seems then
that we should restrict our attention to $r_0'<r_0$. Note that this
also implies that $r_0'<r_0<\re$, so that the interior black hole does
fit inside the enhan\c con shell.

An exceptional case that may be of interest is $r_0' = r_0$. This choice
makes $S_{ij} = 0$ and dramatically simplifies the other components of the
stress tensor.
In particular, the stress--energy takes a relativistic form similar to
that of the BPS configurations, where there are two tensions, one in
the K3 directions and one along the effective membrane directions:
\beqa
\tau_{\rm mem}(r_{\rm i})&=& (N-N')(\hat\tau_6V-\hat\tau_2)
\left({1\over 4\pi r^2_{\rm i} V}\right)\ ,\nonumber\\
\tau_{\rm K3}(r_{\rm i})&=& (N-N')\hat\tau_6\left({1\over 4\pi r^2_{\rm i}}
\right)\ ,\nonumber
\eeqa
where $\hat\tau_6=\sqrt{L(r_{\rm i})}\alpha_6\tau_6$ and
$\hat\tau_2=\sqrt{L(r_{\rm i})}\alpha_2\tau_2$. Hence the results are
essentially the same as in eqn.~\reef{admcold}, but the fundamental
tensions have been modified.  We can model the source as a shell of
wrapped six--branes with a worldvolume action as in eqn.~\reef{efact}
but with modified fundamental tensions. However, one can verify that
these new tensions do not correspond to those of a wrapped D6--brane
for any value of $r_i$. As it appears that there is no six--brane in
string theory with such tensions, the supergravity configurations with
$r_0=r_0'$ would seem to be unphysical.

At this point, we have constrained $r_0'$, but not completely fixed
this free parameter. We leave this issue unresolved here, but we
return to it in the concluding section.  The above discussion should
illustrate for the reader that in general the technique of cutting and
pasting together various supergravity solutions is a crude procedure.
One is not entitled to believe that the resulting configuration is
necessarily physically relevant. Rather, one must supplement this
approach with other physical arguments to ascertain the relevance or
otherwise of the resulting solutions.

To finish this sub--section, we comment that given the discussion of
D6/D2--brane bound states in section 3.2, there is in fact a physical
mechanism by which more D6--branes could accumulate in the black hole.
That is, eqn.~\reef{consvol} determines a critical value $N'_*$ beyond
which a wrapped D6--brane can not enter the interior region. However,
when D2--branes form bound states with the wrapped D6--branes, the
resulting composite branes see no obstruction to entering these
region. In the non--extremal situation, one does not have the freedom to
pull D2--branes in the central black hole out to the enhan\c con shell
to form such bound states which would then fall back in. However, one
can imagine that if these bound states were involved in the initial
infall of branes onto the original Schwarzschild black hole then the
final central black hole could contain any number of D6--branes
between $N'_*\le N'\le N$. Note that since we are considering a
configuration with more D2--branes than D6--branes, \ie with $M>N$,
there would be no obstacle to having the black hole swallow all of the
D6--branes. Hence the bound states provide a mechanism for forming
black holes in which the horizon is surrounded by a region with
$V(r)<V_*$. Furthermore, if the entire black hole moves out from the centre
out to the enhan\c con radius, we imagine that all of the D6--branes
in the shell would be swallowed to form a final state black hole
described by simply the exterior solution
(\ref{junterior}--\ref{juncale}).

\subsection{The Geometry Revisited}
\label{funnydef}

We would like to consider the non--extremal solution in the regime
$M<N$. The limit $M=0$ is of particular interest.  If we take $M<N$ in
the non--extremal solution (\ref{junterior}--\ref{juncale}), the
obvious change is that the scale $r_2$ becomes negative, as can be
seen in eqn.~\reef{juncale}. This simple difference has a drastic
effect on the nature of the solution.  \beq \Z_2=1+{\alpha_2 r_2\over
  r}=\left\lbrace \matrix{1 + {\sqrt{r_2^2+r_0^2/4}-r_0/2\over
      r}&,&\quad r_2>0\ \,\cr 1 - {\sqrt{r_2^2+r_0^2/4}+r_0/2\over
      r}&,&\quad r_2<0\ .\cr} \right.
\label{bore}
\eeq While for $r_2>0$, this harmonic function is always positive for
$r>r_0$, in the case $r_2<0$, $\Z_2$ vanishes at $r=|\alpha_2
r_2|>r_0$.  The latter vanishing results in the appearance of a
repulson--type singularity at this radius. Given our experience with
the BPS configurations, this is precisely the behaviour that we should
might in this regime (in particular, at $M=0$).

We further remark that in solving the supergravity equations there is
in fact a {\it choice} to be made in fixing $\alpha_2$. One finds the
equations are solved by both:
\begin{equation}
\alpha_2 r_2= \pm\sqrt{r_2^2+{r_0^2/4}}-r_0/2\ ,
\label{yourk}
\end{equation}
irrespective of the sign of $r_2$. For our discussion here, the plus
sign is the correct choice for $M>N$, while the minus seems to apply
for $M<N$.  The non--extremal solution as presented in
eqns.~(\ref{junterior}--\ref{juncale}) correctly captures both of these
choices. We return to discuss the sign ambiguity in eqn.~\reef{yourk}
in the final section.

Notice that, crucially, the solution does not make a smooth transition
between $M>N$ and $M<N$.  Consider taking the limit $r_2\rightarrow0$
in eqn.~\reef{bore}. On the top line, it yields $\Z_2=1$, while in the
bottom line, one finds $\Z_2=1-r_0/r$.  (Of course, this discontinuity
vanishes in the BPS limit with $r_0\to 0$.)  This behaviour is
actually problematic. Consider fixing $M$ and taking the limit
$V\to\infty$.  This corresponds to smoothly taking the limit $r_2\to
0$. In this case, our intuition is that the solution should reduce to
that of the usual non--extremal solution for (unwrapped) D6--branes,
as the effective D2--brane sources are being infinitely diluted over
the D6--brane worldvolume. However, we just showed that our solution
for $M<N$ does not properly accomplish this limit. Thus, our
non--extremal solution in this regime seems not to be simply
describing a non--extremal D6--brane wrapped on K3; rather there is
some local difference, independent of the effects of K3 curvature. In
particular these solutions seem to carry nontrivial dilaton hair ---
see concluding remarks.  To find a solution which is locally just a
non--extremal D6--brane, we presumably need to consider a more general
ansatz for the metric and dilaton. In the absence of such a solution,
it is still interesting to apply the techniques we have developed to
the present solution, and we shall in the sequel.

Given the preceding discussion, we regard the current sub--section as an
exploration of a distinct family of supergravity solutions.  Furthermore,
to avoid any confusion about signs, we introduce some new definitions
for the current discussion. For the following, the exterior solution
will taken as given above in eqns.~(\ref{junterior}--\ref{z6ext}),
except that we replace the D2--brane harmonic function with \beq
\Z_2=1 - {\alpha_2 r_2 \over r}\ ,\qquad \alpha_2 = {r_0 \over 2r_2} +
\sqrt{1 + \left( {r_0 \over 2r_2} \right)^2}\ ,
\label{z2ext}
\eeq
where 
\begin{equation}
r_2=r_6{V_*\over V}\left(1-{M\over N}\right)>0\ .
\label{junctale}
\end{equation}
With these definitions, the result for the enhan\c con radius becomes
\begin{equation}
\re = {V_*\alpha_6 r_6 +V \alpha_2 r_2 \over V - V_*}\ .
\label{junktre}
\end{equation}
Note that the D2--branes still cause the enhan\c con radius to shrink,
\ie a larger $M$ yields a smaller $r_2$.

Also note that with $M=0$, this solution corresponds to that given in
ref.~\cite{jpp}, except that we have corrected a sign in $\alpha_2$
relative to that which appears there.  As commented above, in this
regime $M<N$, the supergravity solution acquires a repulson--type
singularity at $r=r_{\rm r}=\alpha_2 r_2$, where the volume to the K3
shrinks to zero. As noted above and as is clear from the definitions
in eqns.~(\ref{junctale}--\ref{junktre}), this repulson radius $r_{\rm
  r}$ is always {\it greater} than the radius $r_0$, at which there
would have been an event horizon. Hence this solution will never yield
a black hole horizon no matter how large $r_0$ becomes.  The absence
of an horizon again suggests that the resulting non--extremal solution
has some additional structure turned on.

Of course, the repulson singularity should be unphysical, as before,
and our task is to determine how to replace this interior geometry for
$r<r_{\rm i}$ (or rather for $r<\re$, as we will argue shortly).  As
our ansatz for the interior geometry, we take precisely the same
interior solution (\ref{tag1}--\ref{tag2}) as presented above.
However, to accommodate our new definitions in eqns.~\reef{z2ext} and
\reef{junctale}, we now write \beq \hH_2 = 1
-{\alpha_2r_2+\alpha_2'r_2'\over r_{\rm i}} +{\alpha_2' r_2' \over r}\ 
.  \eeq Implicitly we assume that the physically relevant interior
still satisfies the relation $N'\le M$ for simplicity. That is, the
black hole that appears in the interior region contains more
D2--branes than wrapped D6--branes.

The results for the shell stress--energy remain as given in
eqn.~\reef{trashij}. Then the only changes come through the implicit
redefinition of the relevant functions considered in this section.

A case of particular interest is $M=0\,(=N')$, \ie no additional D2--branes.
In this case, the harmonic functions in the interior region are simply
constants, and the interior solution reduces to a Schwarzschild black hole
parameterised by $r_0'$. In this case, the stress tensor becomes:
\begin{eqnarray}
2\kappa^2 S_{tt}&=&{1\over\sqrt{G_{rr}}} \left({\Z_2^\prime\over
\Z_2}+{\Z_6^\prime\over \Z_6} +{4\over r_{\rm i}}\left(1-\sqrt{{L\over
K}}\right)\right) G_{tt}\ ,\nonumber\\ 2\kappa^2
S_{\mu\nu}&=&{1\over\sqrt{G_{rr}}} \left({\Z_2^\prime\over
\Z_2}+{\Z_6^\prime\over \Z_6} +{K^\prime\over K} -{L^\prime \over
L}\sqrt{{L\over K}} +{4\over r_{\rm i}}\left(1-\sqrt{{L\over
K}}\right)\right) G_{\mu\nu}\ ,\nonumber\\
2\kappa^2 S_{ij}&=& {1\over \sqrt{G_{rr}}} \left({K^\prime\over
K}-{L^\prime \over L}\sqrt{{L\over K}}+{2\over r_{\rm
i}}\left(1-\sqrt{{L\over K}}\right)\right) G_{ij} \ ,\nonumber\\
2\kappa^2 S_{ab}&=&{1\over \sqrt{G_{rr}}}\left( {\Z_6^\prime\over
\Z_6}+{K^\prime\over K}-{L^\prime \over L}\sqrt{{L\over K}}+{4\over
r_{\rm i}}\left(1-\sqrt{{L\over K}}\right)\right)G_{ab} \ .
\label{warmstressed} 
\end{eqnarray} 

\subsection{Some More Physics} \label{funnyfizz}

As already noted above, the non--extremal solutions presented in the
previous sub--section exhibit some peculiar features. Hence we have little
physical intuition to guide us here, in particular to determine the
interior solution given a fixed set of parameters, $r_0$, $M$ and $N$.
Once again, we have three free parameters to determine $r_0'$, $N'$
and $r_{\rm i}$. To gain some insight, we begin by studying the
exterior solution with two types of probe.

We may probe the geometry with a single wrapped D6--brane.  In this
non--extremal situation there appears a non--trivial potential for the
probe's motion, since we have broken supersymmetry. The result is:
\begin{eqnarray}
{\cal L}&=&T(r,\theta,\phi)-U(r)\ ,\qquad{\rm with}\nonumber\\
\nonumber\\
{T}
&=&{\mu_2 \Z_6(r)\over 2g_s\sqrt{K(r)}}\left[ {{\widehat
V}(r)\over V_*}-1\right] \left[{{\dot r}^2\over K(r)}+r^2({\dot
\theta}^2+\sin^2\theta{\dot\phi}^2)\right]\label{tagger}\\
{U}&=&{\mu_2\over g_s
\Z_2(r)}\left[\left({{\widehat V}(r)\over V_*}-1\right)\sqrt{K(r)}
+{1\over\alpha_2}-{1\over\alpha_6}{{\widehat V}(r)\over V_*}\right]\ .
\label{lagrange}
\end{eqnarray}
From eqn.~\reef{tagger}, we see that the effective membrane tension
vanishes precisely at the enhan\c con radius where
$\widehat{V}(r)=V_*$, as before. Furthermore, it becomes negative for
smaller radii where $\widehat{V}(r)<V_*$.  It is easy to see that the
potential in eqn.~\reef{lagrange} is always attractive. In fact, it is
increasingly attractive for larger $r_0$.  Given these results, it
seems that the physically relevant solution will be that with $r_{\rm
  i}=\re$, as there is no obstacle to preventing the infall of the
wrapped D6--branes at larger radii, and as before the exterior solution
yields unphysical effects for $r<\re$.

\begin{figure}[ht]
\centerline{\psfig{figure=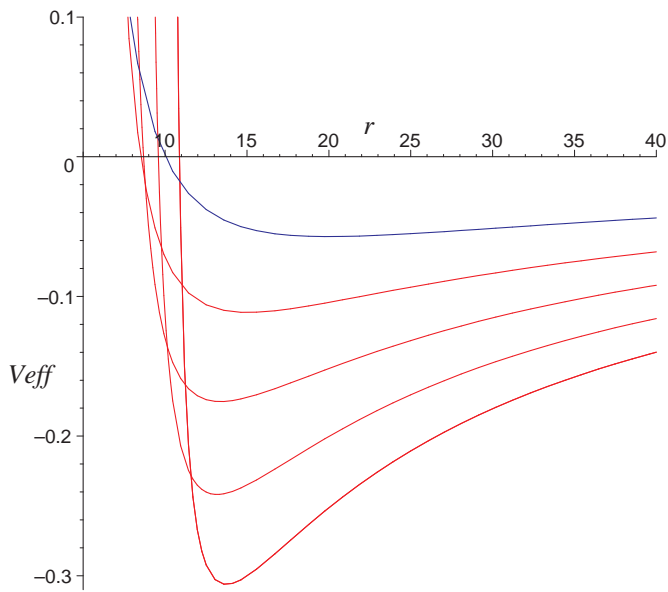,height=4.0in}}
\caption{\small The effective potential for increasing amounts of
non--extremality, $r_0$. The top curve is the extremal case.}
\label{veff}
\end{figure}

\begin{figure}[ht]
\centerline{\psfig{figure=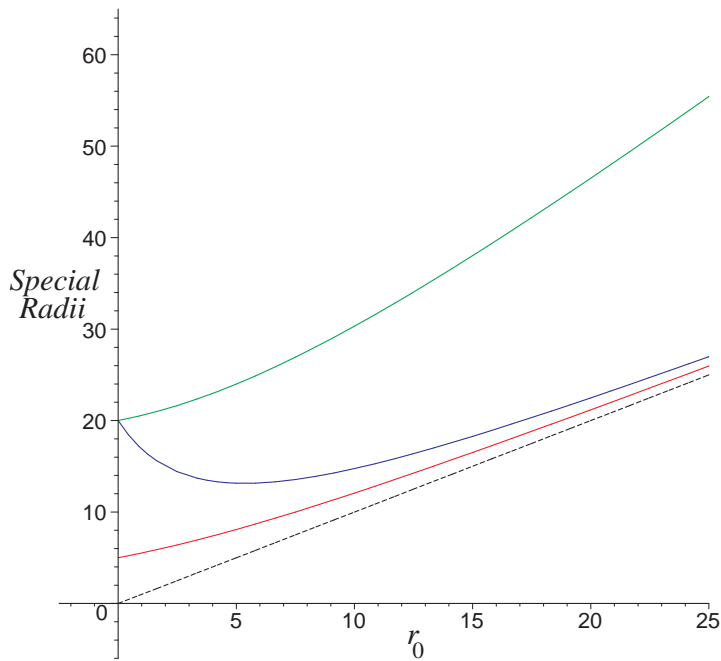,height=4.0in}}
\caption{\small Some special radii for increasing non--extremality,
  $r_0$, in the case $M=0$. The {\it solid} curves are as follows: The
  enhan\c con radius is top, the repulson radius is bottom, and the
  minimal radius for which excision would remove all repulsiveness,
  ${\hat r}_{\rm d}$, (which also deserves a name) is the middle
  curve. Notice that the latter two  coalesce into the would--be horizon
  (dotted  line at very bottom) at large $r_0$.}
\label{radii}
\end{figure}

We may also probe the geometry with point particles as we did for the
extremal geometry.  The result is identical to that given in
(\ref{radial}), and is shown in figure~\ref{veff}.  Again we find that
the geometry is purely attractive up to the radius ${\hat r}_{\rm d}$
determined by the vanishing of the first derivative of the string
frame component $G_{rr}$:
\begin{equation}
 {\partial\over\partial
r}\left(\Z_2(r)\Z_6(r)K^{-2}(r)\right)\biggr|_{r={\hat r}_{\rm d}}=0\
, \quad {\rm where} \quad {\hat r}_{\rm d}= r_6\left[
{\alpha_6Vr_0-\alpha_2 V_*(1-M/N)(r_0+2 r_6 \alpha_6)\over
\alpha_2V_*r_6(1-M/N)-
V(2r_0+ r_6 \alpha_6) }\right]\ .
\label{crucialthree}
\end{equation}
This is the place where the repulsive part of the geometry begins.
The special radii ${\hat r}_{\rm e}, {\hat r}_{\rm d}$ are not the
same away from extremality.  The radius ${\hat r}_{\rm d}$ is always
{\it less than} the enhan\c con radius ${\hat r}_{\rm e}$. Hence
introducing the shell at $r_{\rm i}=\re$ will again ensure that all of
the repulsive behaviour in the metric is removed.  Figure~\ref{radii}
shows the behaviour of these special radii for increasing $r_0$, for
some generic choice of the parameters, with $M=0$.

The remaining discussion precisely follows that in section 4.2. First,
we would say that (unless $M=0$) not all of the D6--branes will stay
at the enhan\c con radius. Rather they will fall into the central
black hole until the K3 volume is fixed at $V_*$ over this entire region.
This condition is determined by precisely the same equation as before,
\ie eqn.~\reef{consvol}, which implicitly yields precisely the same
constraint as before, $N'=N_*'(r_0)$. So as before, the physical solution
would place $N'=N_*'\simle M/2$ wrapped D6--branes on the central black
hole.

Furthermore our discussion of $r_0'$ is also as in section 4.2. By
examining the transverse stresses in the shell, we argue that we must
have $r_0'<r_0$. However, note that in this case, there was no
restriction on $r_0$, and so we must impose $r_0'<\re$ as a separate
constraint, which ensures that the central black hole actually fits
inside of the enhan\c con shell. Beyond these two constraints, we have
not fixed the final free parameter, $r_0'$.

\section{Concluding Remarks}
\label{concl}

We have found that a purely supergravity analysis singles out the
enhan\c con radius as a special place in spacetime. For the BPS
enhan\c con, our precise matching calculation was able to verify that
the solution proposed in ref.~\cite{jpp} corresponds precisely to
introducing a shell of wrapped D6--branes at the enhan\c con radius.
This result shows that the excision procedure suggested by the full
string theory is also a sensible construction from the point of view
of supergravity. Our analysis also strengthens the case that there is
simply a flat--space region inside the enhan\c con shell (and not,
{\it e.g.}, some breakdown of the spacetime description).

Furthermore, these matching calculations explain why one can sensibly
carry out the excision procedure in supergravity, even without full
knowledge of the detailed mechanism by which the branes will expand.
The latter is especially pertinent to wrapped D$p$--branes with $p\not=6$,
where one does not have the intuition supplied by the fact the the
expansion is just the physics of ordinary BPS Yang--Mills--Higgs
monopoles, and also for uplifts of the enhan\c con to M--theory, as
studied in ref.~\cite{jj}.

It is clear that this technique has wide applications.
Some cases of particular interest are Denef's ``empty holes'' studied
in ref.\cite{denef}.  There, a similar excision procedure is
performed, matching a non--trivial supergravity exterior to a flat
space interior.  This matching seems to be dictated by supersymmetry
and the attractor flow equations. However, the continuity conditions
imposed at the excision surface guarantee that the implicit source
consists of a shell of massless particles as would be appropriate for
a D3--brane wrapping a conifold cycle. It may be interesting to study
those solutions from this point of view in more detail.

We used the supergravity analysis to explore a non--extremal
deformation of the enhan\c con\footnote{Refs.~\cite{berglund} provide
other opportunities to apply the techniques discussed here to
non-supersymmetric cases}. However, in this case, our results seemed
less satisfactory, as we were unable to completely fix $r_0'$, the
radius of the black hole in the interior region. Instead we were only
able to produce two relatively lax constraints: $r_0'<r_0$ and
$r_0'<\re$.

In fact, we expect that there is no single correct value for $r_0'$ in
the following sense: The non--extremal configurations consist of two
thermal subsystems: the central black hole with the Hawking
temperature, and the enhancon shell whose internal degrees of freedom
are thermally excited. (We could extend the number of subsystems to
three by including a thermal bath at infinity.) While the solution
presumably describes these subsystems in thermal equilibrium for one
particular value of $r_0'$, we are implicitly working in a regime
where they are only weakly coupled, \eg thermal fluxes are diffuse
enough that their gravitational back reaction is negligible. Hence if
the black hole and the enhan\c con shell have different temperatures,
we expect that they will only equilibrate over a very long time
span. Hence these static non--extremal solutions can be considered as
good approximations to the full system which only evolves very
slowly. This would be analogous to how isolated black holes are
described by classical solutions of Einstein's equations and the
dynamical effects of Hawking evaporation are ignored. With this
reasoning, it is natural to think that $r_0'$ should remain a free
parameter (subject to certain broad constraints) in the non--extremal
solutions.

Our partial results with respect to $r_0'$ provide a good illustration of
the fact that the excision technique of matching together different
supergravity solutions is a coarse tool when there is no
supersymmetry. To completely verify the physical correctness of the
results, one must still invoke other arguments. In particular,
if the non--extremal solutions are to describe thermally excited
enhan\c cons, it seems that we need a better microscopic understanding
of the thermal physics of the D--branes in the shell.  While it is
clear that the R--R sources remain unchanged, \ie the R--R fields are
simply determined by the number of branes, the thermal excitation of
the internal modes on the branes will modify the metric and dilaton
sources. While the matching calculations allow us to calculate these
sources from the supergravity solution for a particular choice of
parameters, without a microscopic model we cannot verify, \eg what
the associated temperature should be. In principle, such a microscopic
picture would allow us to identify the value of $r_0'$ at which the
central black hole and the enhan\c con shell would be in thermal
equilibrium.

In fact, a larger problem was revealed in examining the non--extremal
solutions in the regime where the D6--branes outnumbered the
D2--branes. In this case, the solutions exhibit a number of peculiar
features. First, the exterior solution never contains an event horizon
no matter how large the non--extremality scale $r_0$ became.
Furthermore, in limit $r_2\rightarrow0$, equivalent to a large K3 volume
limit, these solutions do not seem to reproduce the expected
non--extremal D6--brane solution. Examining this limit more closely, we
see that the R--R 3--form potential actually vanishes at $r_2=0$
even though we have a nontrivial harmonic function $\hH_2=1-r_0/r$. On
the other hand, this harmonic function still makes a nontrivial
contribution to the metric, and in particular to the dilaton. Hence,
rather than the expected solution, we seem to have produced a solution
for non--extremal D6--branes carrying some additional dilaton hair. In
fact (as in the extremal case), the repulson singularity in these
solutions implies that the usual no--hair theorems no longer apply, and
so our non--extremal solutions are presumably just one example of a
family of singular solutions with the same asymptotic charges and
mass, but differing by scalar hair.  It would be interesting if these
generalised solutions could be found, along the lines of
ref.~\cite{fern}, as some of the physical problems might be
ameliorated with the appropriate hair dressing. However, without the
guidance of supersymmetry in these non--extremal solutions, this seems a
daunting task.

\begin{figure}[ht]
\centerline{\psfig{figure=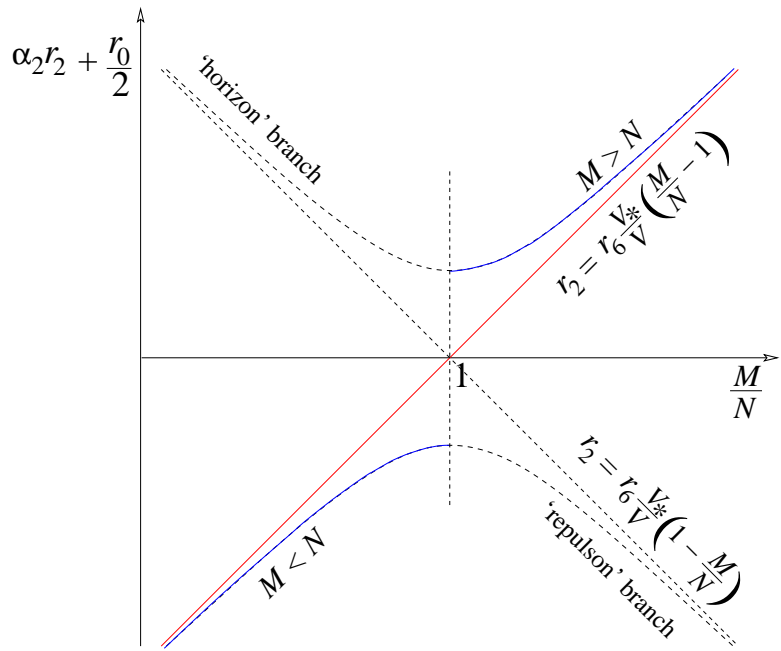,height=3.5in}}
\caption{\small 
  The two branches (solid lines) of supergravity solutions for varying
  values of the ratio $M/N$. There is a crucial discontinuity at
  $N=M$. (See text.)}
\label{branches}
\end{figure}

Finally, recall our observation that solving the supergravity equations 
left a sign ambiguity in the following relation:
\begin{equation}
\alpha_2 r_2= \pm\sqrt{r_2^2+{r_0^2/4}}-r_0/2\ ,
\label{yourrk}
\end{equation}
irrespective of the sign of $r_2$. In the discussion in section 4,
we made the particular choices that the sign is plus (minus)
for $M>N$ ($M<N$). Figure~\ref{branches} illustrates the two different
branches in eqn.~\reef{yourrk} and our choices in resolving the
ambiguity. A feature that distinguishes our choice of signs is that
in the limit $r_0\to 0$, these two branches coalesce to fill out the
diagonal which we know corresponds to the proper BPS solution. However,
as illustrated, the two families of solutions actually both extend over
the entire range of $M/N$. The solutions on the lower branch are all
characterised by having a repulson singularity (before any excision
procedure is implemented). As discussed above, these singular solutions
are not unique and we will not consider them further here.

A distinguishing feature of the solutions corresponding to the upper
branch is that all of these non--extremal configurations are black
holes (again before any excision procedure is implemented). Hence we
expect that no--hair theorems will single out these solutions as the
unique solutions with a nonsingular event horizon for a given set of
parameters: $r_0$, $M$ and~$N$.  Therefore, it is to be expected that
there is some interesting physics associated with these solutions even
when $M<N$.  For example, if we probe a ``large'' Schwarzschild black
hole with either test D6-- or D2--brane probes, there are now
obstruction to either type of probe from falling into the black hole.
So if this black hole absorbs a relatively small number of branes,
we expect that the end state must be a black hole in this family of
solutions, even if $M<N$. Similarly, one might think about starting
with a black hole on the $M>N$ part of this branch and dropping in
anti--D2--branes to reduce the final black hole to $M<N$.  As a final
comment, however, we note that even with small $r_0$ the black holes
with $M<N$ have masses (relatively) far above the BPS limit.

\bigskip
\noindent{\bf\Large Acknowledgements}

\noindent Research by RCM was supported by NSERC of Canada and Fonds
FCAR du Qu\'ebec. Research by AWP was supported by NSERC of Canada and
the University of Toronto. Research by CVJ and SFR was partially
supported by PPARC of the United Kingdom.  We would like to thank the
Aspen Center for Physics for hospitality during the initial stages of
this project.  CVJ would like to thank the ITP at Stanford for
hospitality, during the latter stages of this project. AWP would like
to thank Emil Martinec, Sumati Surya and Lenny Susskind for useful
discussions.  RCM would like to thank Neil Constable and Frederik
Denef for useful discussions. We thank Joe Polchinski for very useful
conversations. CVJ, AWP and SFR would like to thank the ITP in Santa
Barbara for hospitality, and the organisers of the {\it M--theory}
workshop for the opportunity to participate. We thank all of
participants of the workshop for remarks.  Research at the ITP (Santa
Barbara) was supported in part by the U.S.  National Science
Foundation under Grant No. PHY99--07949.

\end{document}